\documentclass{article}%
\usepackage{amsfonts}
\usepackage{amssymb}
\usepackage{amsmath}
\usepackage{graphicx}%
\providecommand{\U}[1]{\protect\rule{.1in}{.1in}}

\begin{document}

\title{Generative Mechanisms:\\The mechanisms that implement codes}
\author{David Ellerman\\University of Ljubljana}
\maketitle

\begin{abstract}
\noindent The purpose of this paper is to abstractly describe the notion of a
generative mechanism that implements a code and to provide a number of
examples including the DNA-RNA machinery that implements the genetic code,
Chomsky's Principles \& Parameters model of a child acquiring a specific
grammar given `chunks' of linguistic experience (which play the role of the
received code), and embryonic development where positional information in the
developing embryo plays the role of the received code. A generative mechanism
is distinguished from a selectionist mechanism that has heretofore played an
important role in biological modeling (e.g., Darwinian evolution and the
immune system).

\end{abstract}

Keywords: generative mechanisms, selectionist mechanisms, genetic code,
Chomsky Principles \& Parameters, embryonic development, information theory,
logical entropy and Shannon entropy.
\title{Generative Mechanisms:\\The mechanisms that implement codes}
\author{David Ellerman\\University of Ljubljana}
\maketitle

\begin{abstract}
\noindent The purpose of this paper is to abstractly describe the notion of a
generative mechanism that implements a code and to provide a number of
examples including the DNA-RNA machinery that implements the genetic code,
Chomsky's Principles \& Parameters model of a child acquiring a specific
grammar given `chunks' of linguistic experience (which play the role of the
received code), and embryonic development where positional information in the
developing embryo plays the role of the received code. A generative mechanism
is distinguished from a selectionist mechanism that has heretofore played an
important role in biological modeling (e.g., Darwinian evolution and the
immune system).

\end{abstract}
\tableofcontents

\section{Introduction}

There are a number of mechanisms which can be abstractly defined and are found
to be implemented in biology. The best-known is the selectionist mechanism of
biological evolution. But that mechanism can be abstractly described and then
implemented in non-biological settings as in `evolutionary' or selectionist
programs on computers. In that mechanism, a wide variety of different options
are randomly generated and then whittled down to a `select few' using some
fitness criterion.

Our purpose in this paper is to abstractly describe another type of mechanism,
a generative mechanism, that is also found to be implemented in biology. A
generative mechanism can be abstractly described using the graph-theoretic
notion of a tree usually pictured upside down with the root at the top and
then the branches going downward to eventually terminate in the leaves. 

\textbf{Definition}: A \textit{generative mechanism} implements a
\textquotedblleft code\textquotedblright\ that determines which branch of a
tree is taken as one descends from the root to a specific leaf that was
encoded in the code.%

\begin{center}
\includegraphics[
height=1.5851in,
width=2.9769in
]%
{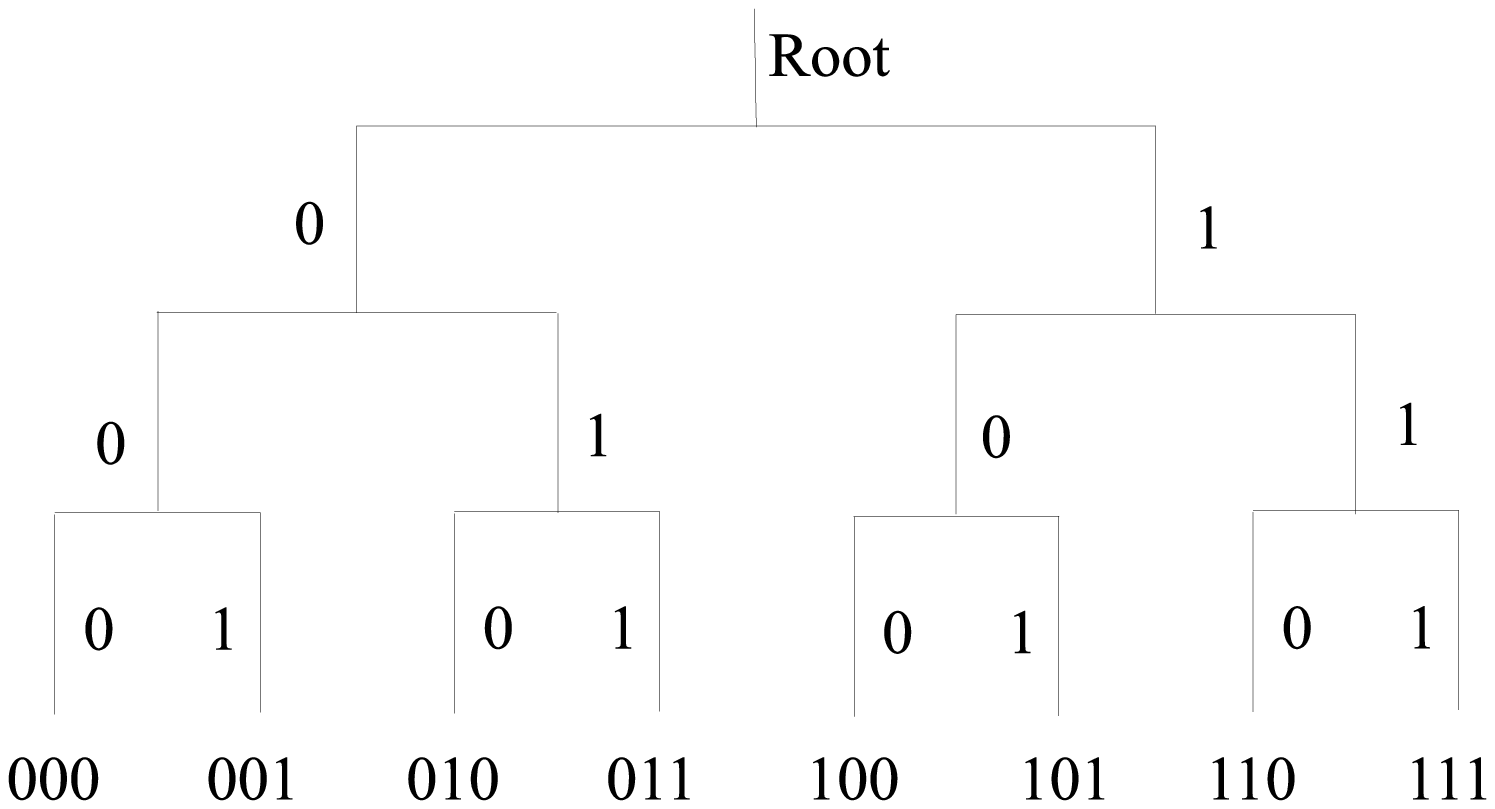}%
\end{center}

\begin{center}
Figure 1: Generative mechanisms can be illustrated by tree diagrams.
\end{center}

For instance, in Figure 1, reading the three-letter binary codes words from
left to right, the code word $011$ is implemented by taking the $0$-branch at
the first branching point and then the $1$-branch at the next two branching
points.\footnote{The use of a binary code for an illustration does not imply
that generative mechanisms are limited to binary codes, e.g., the genetic code
has a code alphabet of four letters.} 

One can think of a code as a hierarchical set of switches. The number of
settings on each switch (aside from neutral) is the number of letters in the
code alphabet. Each switch determines a branching point in the tree as in
Figure 2.
\begin{center}
\includegraphics[
height=2.134in,
width=3.6725in
]%
{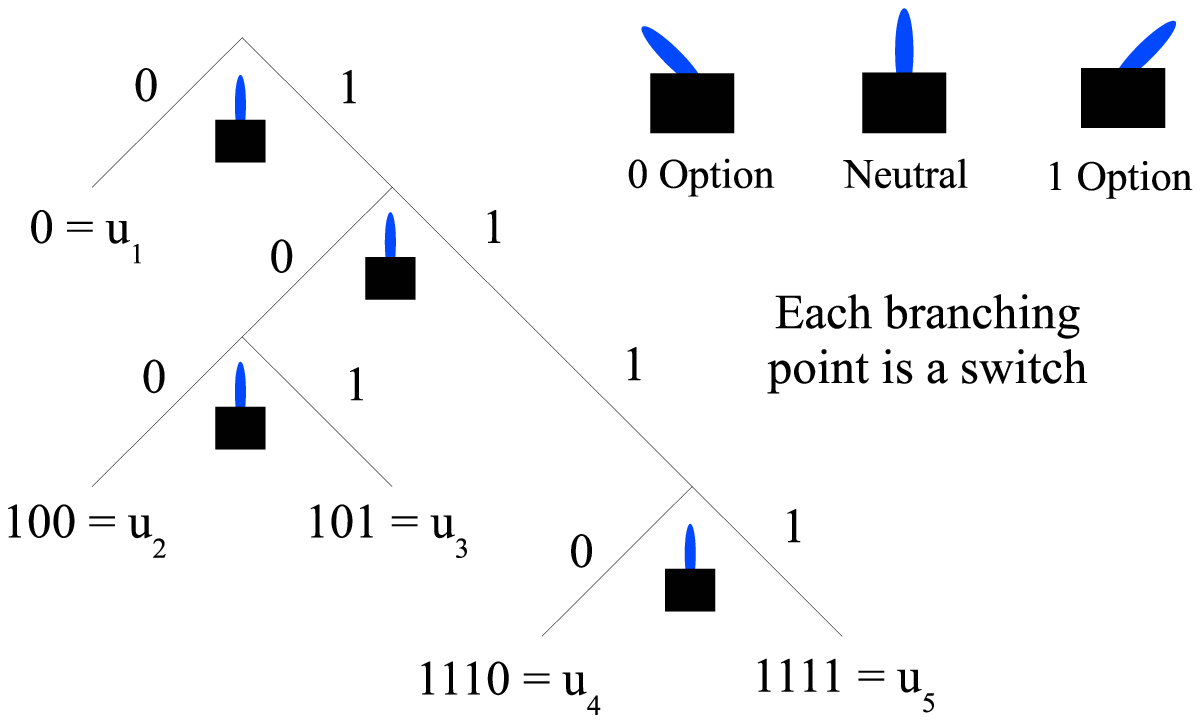}%
\end{center}

\begin{center}
Figure 2: Generative mechanism specified by a hierarchy of switches.
\end{center}

An everyday example of a generative mechanism is the game of 20-questions
where a player tries to traverse an implicit tree of binary branching points
(yes-or-no questions) to determine the hidden answer at a leaf of the tree. By
implementing a code, a generative mechanism navigates down through a diverse
set of possible outcomes, represented by the leaves on the tree, to reach a
specific outcome or message. 

\section{Partitions and codes}

Mathematically, a partition on a set represents one way to differentiate the
elements of the set into different blocks. The join with another partition
generates a partition with more refined (smaller) blocks that makes all the
distinctions of the partitions in the join. Starting from a single block
consisting of the set of all possibilities (like the unbranched root of the
tree), a sequence of partitions joined together differentiates all the
elements of the set ultimately into singleton blocks that are the leaves of
the tree. All the (prefix-free) codes of coding theory can be generated in
this way and then the codes are implemented in practice to generate the coded
outcomes. 

A \textit{partition} $\pi=\left\{  B_{1},...,B_{m}\right\}  $ is a set of
non-empty subsets $B_{i}$, called \textit{block}s, of a universe set $U$ that
are disjoint and whose union is all of $U$---as in Figure 3.%

\begin{center}
\includegraphics[
height=1.641in,
width=1.9595in
]%
{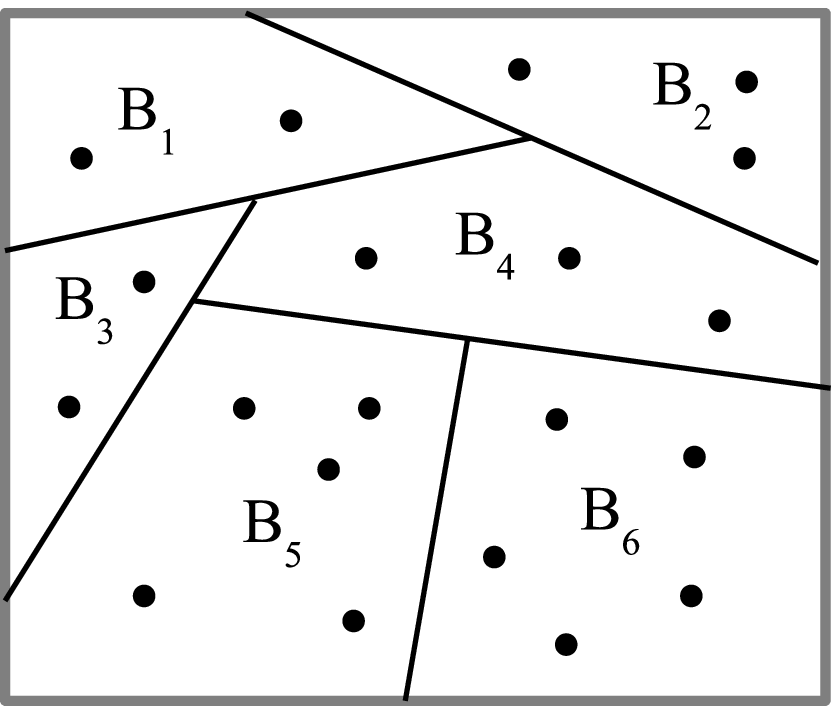}%
\end{center}

\begin{center}
Figure 3: A partition with six blocks on a universe set $U$.
\end{center}

The general idea of a partition is that the elements in the same block are
considered indistinct from one another while the elements in different blocks
are considered as distinct. Given another partition $\sigma=\left\{
C_{1},...C_{n}\right\}  $, the join $\pi\vee\sigma$ of the two partitions is
the partition whose blocks are the non-empty intersections $B_{i}\cap C_{j}$
of the blocks of the two partitions.
\begin{center}
\includegraphics[
height=1.5817in,
width=6.3512in
]%
{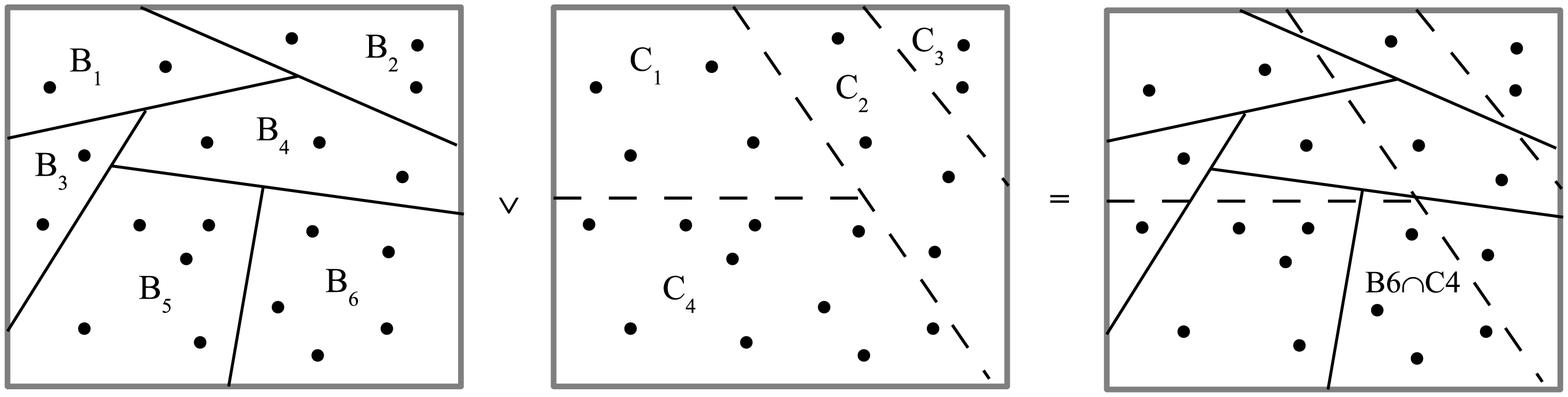}%
\end{center}

\begin{center}
Figure 4: Join of $\pi$ and $\sigma$ = partition of non-empty intersections
such as $B_{6}\cap C_{4}$.
\end{center}

With consecutive joins of partitions (always on the same universe set), the
blocks get smaller and smaller until they reach the discrete partition
$\mathbf{1}_{U}$ with the smallest non-empty blocks with are the singletons of
elements of $U$. The least refined partition is the indiscrete partition
$0_{U}=\left\{  U\right\}  $ whose only block is all of $U$ and it represents
the root of the tree that would illustrate the consecutive joins in Table 1.
The indiscrete partition is the identity for the join operation: $0_{U}\vee
\pi=\pi$ for any partition $\pi$.

\begin{center}%
\begin{tabular}
[c]{|c|c|c|}\hline
Partitions & Consecutive Joins (tree) & Codes\\\hline\hline
$\{\{u_{1},u_{2},u_{3},u_{4},u_{5}\}\}=\mathbf{0}_{U}$ & $\{\{u_{1}%
,u_{2},u_{3},u_{4},u_{5}\}\}=\mathbf{0}_{U}$ & \\\hline
$\{\{u_{1}\},\{u_{2},u_{3},u_{4},u_{5}\}\}$ & $\{\{u_{1}\},\{u_{2},u_{3}%
,u_{4},u_{5}\}\}$ & $0=$ (code for) $u_{1}$\\\hline
$\{\{u_{1},u_{2},u_{3}\},\{u_{4},u_{5}\}\}$ & $\{\{u_{1}\},\{u_{2}%
,u_{3}\},\{u_{4},u_{5}\}\}$ & \\\hline
$\{\{u_{1},u_{2}\},\{u_{3},u_{4},u_{5}\}\}$ & $\{\{u_{1}\},\{u_{2}%
\},\{u_{3}\},\{u_{4},u_{5}\}\}$ & $100=u_{2},101=u_{3}$\\\hline
$\{\{u_{1},u_{2},u_{3},u_{4}\},\{u_{5}\}\}$ & $\{\{u_{1}\},\{u_{2}%
\},\{u_{3}\},\{u_{4}\},\{u_{5}\}\}=\mathbf{1}_{U}$ & $1110=u_{4},1111=u_{5}%
$\\\hline
\end{tabular}

Table 1: Prefix-free codes generated by consecutive partition joins.
\end{center}

A code is called \textit{prefix-free} or \textit{instantaneous} if no code
word is the beginning of another code word. All prefix-free code words can be
obtained by a sequence of consecutive partition joins in the following manner.
The number of letters in the code alphabet is the number of blocks in each
partition. All the partitions in the Partitions column of Table 1 (except the
indiscrete partition representing the root) have two blocks with the left
blocks associated with $0$ and the right block associated with $1$. When
taking consecutive joins, once a singleton block appears in the Consecutive
Joins column, it stays a singleton since it cannot be split any further. 

The code for each $u_{i}$ in $U$ is generated in the Partitions column by the
sequence of blocks in the binary partitions (ignoring $\mathbf{0}_{U}$)
containing the element $u_{i}$, with each block contributing a $0$ or $1$ to
the code word for the element until it appears in a singleton block in the
Consecutive Joins column of Table 1. Then the code word stops so that code
word cannot be the prefix for the code word for any other element of $U$. For
instance, consider the element $u_{2}$ which appears in the $1$-block in the
first partition $\{\{u_{1}\},\{u_{2},u_{3},u_{4},u_{5}\}\}$ and then in the
$0$-block in the next two partitions (in the Partitions column of the table).
The element $u_{2}$ first becomes a singleton in the third row join so tracing
its history through the blocks generates the code $100=u_{2}$. 

Corresponding to the generation of a code by consecutive partition joins as in
Table 1, one can construct a tree diagram where each branching is accomplished
by a partition join and each leaf corresponds to when an element first appears
in a singleton. The tree corresponding to Table 1 is the tree in Figure 2.

\section{The genetic code}

The most famous code is, of course, the genetic code which is prefix-free so
it can be generated by a sequence of partitions. In this case, each partition
has four blocks corresponding to the four code letters U, C, A, and G in the
code alphabet. For the partitions in Figure 5, which correspond to the
(non-indiscrete) partitions in the Partitions column in Table 1, the
consecutive joins give all 64 singletons after three branchings or joins so
the amino acids have 3-letter code words. The code is redundant since there
can be several codes for the same acid.

The circles in Figure 5 trace out the code for Thr4 (one of the code words for
Thr, Threonine) which is ACG = Thr4. Note that the order of the partitions
counts in the consecutive-joins determination of the genetic codes. A
different ordering gives a different code which may not describe the operation
of the DNA-RNA machinery to produce a certain amino acid from a given code word.%

\begin{center}
\includegraphics[
height=2.2475in,
width=3.1362in
]%
{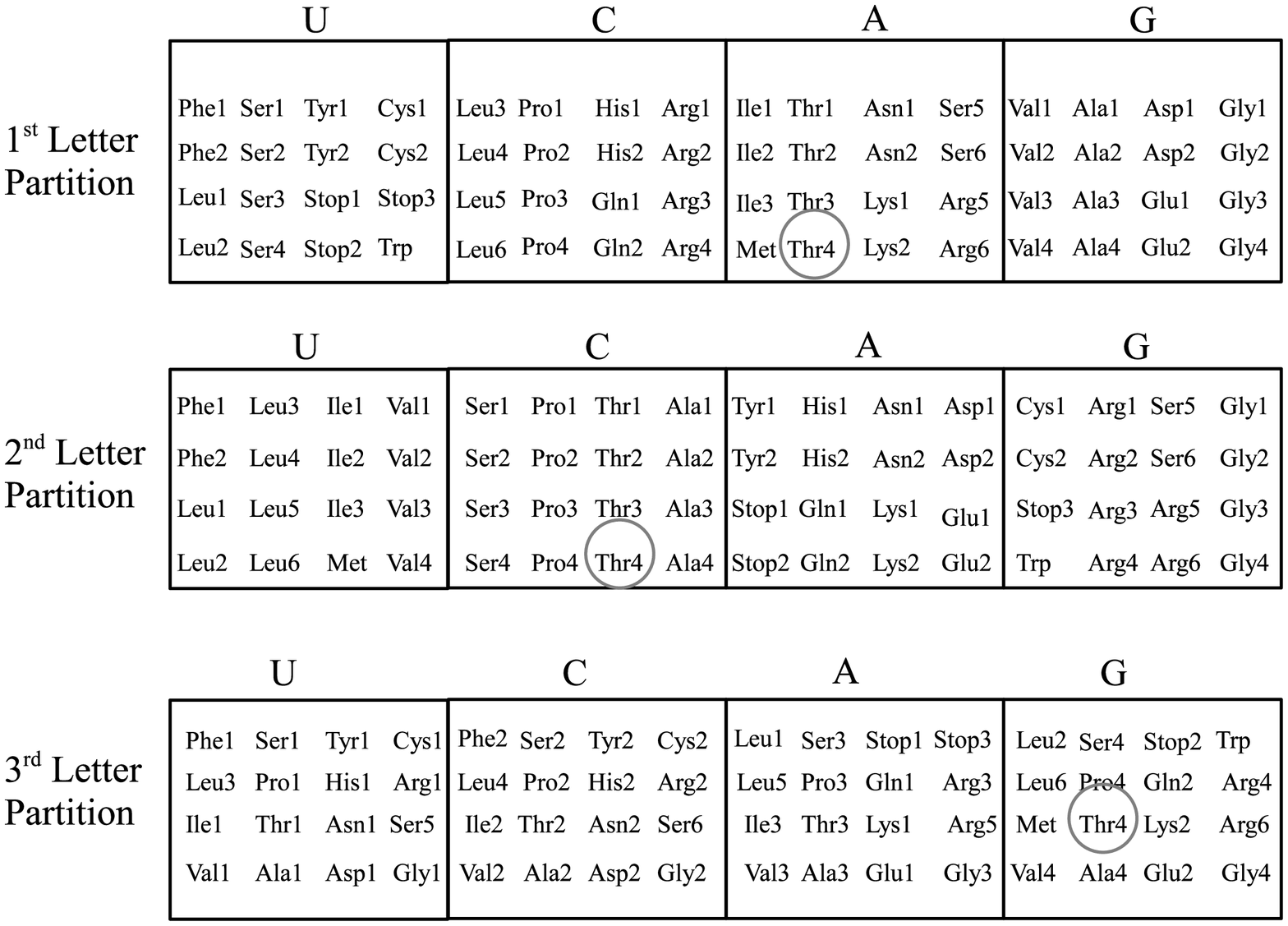}%
\end{center}

\begin{center}
Figure 5: The three partitions that generate the genetic code.
\end{center}

In terms of a tree diagram, the tree would branch four ways at each branching
point and there are three levels, so there are $4^{3}=64$ leaves in the tree.
The generative mechanism associated with the genetic code is the whole DNA-RNA
machinery that generates the amino acid as the output from the code word as
the input. If we abstractly represent the DNA-RNA machinery as that tree with
$64$ leaves, then the given code word tells the machinery how to traverse the
tree to arrive at the desired leaf.

\section{The Principles \& Parameters mechanism for language acquisition}

Noam Chomsky's Principles \& Parameters (P\&P) mechanism
(\cite{chomp-lasnik:pandp}; \cite{chomsky:minimalist}) for language learning
can be modeled as a generative mechanism. Again, we can consider a tree
diagram where each branching point has a two-way switch to determine one
grammatical rule or another in the language being acquired.

\begin{quotation}
\noindent A simple image may help to convey how such a theory might work.
Imagine that a grammar is selected (apart from the meanings of individual
words) by setting a small number of switches - 20, say - either "On" or "Off."
Linguistic information available to the child determines how these switches
are to be set. In that case, a huge number of different grammars (here, 2 to
the twentieth power) will be prelinguistically available, although a small
amount of experience may suffice to fix one. \cite[p. 154]{higginbotham}
\end{quotation}

And the reference to \textquotedblleft20\textquotedblright\ recalls the game
of \textquotedblleft20 questions\textquotedblright\ where the answers to the
questions guides one closer and closer to the desired hidden answer. Chomsky
uses the Higginbotham model to describe a Universal Grammar (UG) as a
generative mechanism.

\begin{quotation}
\noindent Many of these principles are associated with parameters that must be
fixed by experience. The parameters must have the property that they can be
fixed by quite simple evidence, because this is what is available to the
child; the value of the head parameter, for example, can be determined from
such sentences as John saw Bill (versus John Bill saw). Once the values of the
parameters are set, the whole system is operative. Borrowing an image
suggested by James Higginbotham, we may think of UG as an intricately
structured system, but one that is only partially \textquotedblright wired
up.\textquotedblright\ The system is associated with a finite set of switches,
each of which has a finite number of positions (perhaps two). Experience is
required to set the switches. When they are set, the system functions. The
transition from the initial state S0 to the steady state Ss is a matter of
setting the switches. \cite[p. 146]{chomsky:knowoflang}
\end{quotation}

In the tree modeling of the P\&P approach, the relative poverty of linguistic
experience that sets the switches plays the role of the code that guides the
mechanism from the undifferentiated root state (all switches at neutral) to
the final specific grammar represented as a leaf.

The question about the acquisition of a grammar is a good topic to compare and
contrast a selectionist mechanism with a generative mechanism. What would a
selectionist approach to learning a grammar look like? A child would (perhaps
randomly) generate a diverse range of babblings, some of which would be
differentially reinforced by the linguistic environment (e.g.,
\cite{skinner:behave}).

\begin{quotation}
\noindent Skinner, for example, was very explicit about it. He pointed out,
and he was right, that the logic of radical behaviorism was about the same as
the logic of a pure form of selectionism that no serious biologist could pay
attention to, but which is [a form of] popular biology -- selection takes any
path. And parts of it get put in behaviorist terms: the right paths get
reinforced and extended, and so on. It's like a sixth grade version of the
theory of evolution. It can't possibly be right. But he was correct in
pointing out that the logic of behaviorism is like that [of na\"{\i}ve
adaptationism], as did Quine. \cite[Section 10]{chomp-mcgil:interview}
\end{quotation}

As noted, Willard Quine adopts essentially the behaviorist/selectionist account.

\begin{quotation}
\noindent An oddity of our garrulous species is the babbling period of late
infancy. This random vocal behavior affords parents continual opportunities
for reinforcing such chance utterances as they see fit; and so the rudiments
of speech are handed down. \cite[p. 73]{quine:wando}

It remains clear in any event that the child's early learning of a verbal
response depends on society's reinforcement of the response in association
with the stimulations that merit the response, from society's point of view,
and society's discouragement of it otherwise. \cite[p. 75]{quine:wando}
\end{quotation}

A more sophisticated version of a selectionist model for the
language-acquisition faculty or universal grammar (UG) could be called the
format-selection (FS) approach (Chomsky, private communication). The diverse
variants that are actualized in the mental mechanism are different sets of
rules or grammars. Then given some linguistic input from the linguistic
environment, the grammars are evaluated according to some evaluation metric,
and the best rules are selected.

\begin{quotation}
\noindent Universal grammar, in turn, contains a rule system that generates a
set (or a search space) of grammars, \{G1, G2,\ldots, Gn\}. These grammars can
be constructed by the language learner as potential candidates for the grammar
that needs to be learned. The learner cannot end up with a grammar that is not
part of this search space. In this sense, UG contains the possibility to learn
all human languages (and many more). ... The learner has a mechanism to
evaluate input sentences and to choose one of the candidate grammars that are
contained in his search space. \cite[p. 292]{nowak-komarove}
\end{quotation}

After a sufficient stream of linguistic inputs, the mechanism should converge
to the best grammar that matches the linguistic environment. Since it is
optimizing over sets of rules, this model at least takes seriously the need to
account for the choice of rules (rather than just assuming the child can infer
the rules from raw linguistic data). Early work (through the 1970s) on
accounting for the language-acquisition faculty or universal grammar (UG)
seems to have assumed such an approach. The problems that eventually arose
with the FS approach could be seen as the conflict between descriptive and
explanatory adequacy. 

\begin{quotation}
\noindent It was an intuitively obvious way to conceive of acquisition at the
time for---among other things---it did appear to yield answers and was at
least more computationally tractable than what was offered in structural
linguistics, where the alternatives found in structural linguistics could not
even explain how that child managed to get anything like a morpheme out of
data. But the space of choices remained far too large; the approach was
theoretically implementable, but completely unfeasible. \cite[Appendix
IX]{chomp-mcgil:interview}
\end{quotation}

In order to describe the enormous range of human language grammars, the range
of grammars considered would make for an unfeasible computational load of
evaluating the linguistic experience. If the range was restricted to make
computation more feasible, then it would not explain the variety of human
languages. Hence the claim is that the P\&P generative mechanism gives a more
plausible account of human language acquisition than a behavioral/selectionist
approach.\footnote{For more of the mathematical background, see
\cite{ell:4ways}) and the references therein.}

\section{Embryonic development}

The role of stem cells in the development of an embryo into a full organism
can again be modeled as a generative mechanism. The original fertilized egg
becomes the stem cell that is the root of the tree. As illustrated in Figure
6, stem cells come in three general varieties: A) the stem cells that can
reproduce undifferentiated copies of themselves, B) the stem cells that can
reproduce but can also produce a somewhat differentiated cell, and C) a
specialized differentiated cell. Each branching point in a tree has a certain
number of possible leaves or terminal types of cells beneath it in the tree.
In a division (\#1) of an A-type cell, each of the resulting A-type cell could
have a full set of leaves beneath it. But when it splits (\#2) into another
A-type cell and a B-type cell, then the B-cell has a restricted number of
leaves beneath it. The B-type cells can split (\#3) in two, and finally when a
B-type cell gives rise (\#4) to a specific C-type of cell, that is a terminal
branch, i.e., a leaf, in the tree.
\begin{center}
\includegraphics[
height=3.0388in,
width=1.7443in
]%
{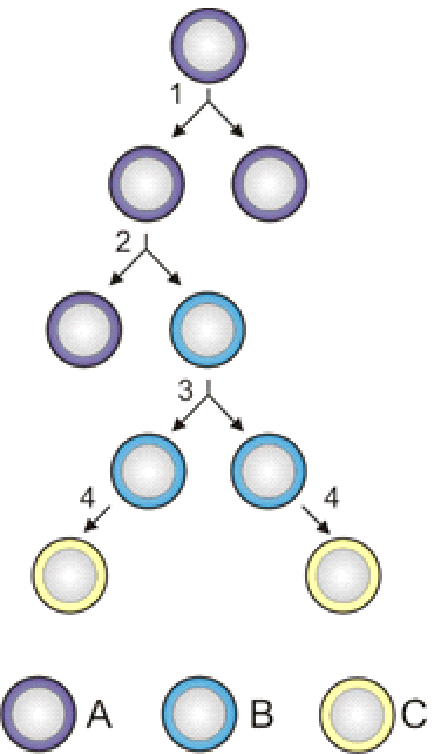}%
\end{center}

\begin{center}
[Attribution: Peter Znamenskiy, CC BY-SA 3.0%

$<$%
http://creativecommons.org/licenses/by-sa/3.0/%
$>$%
, via Wikimedia Commons]

Figure 6: Stem cell division and differentiation A: stem cell; B: progenitor
cell; C: differentiated cell; 

1: symmetric stem cell division; 2: asymmetric stem cell division; 3:
progenitor division; 4: terminal differentiation.
\end{center}

The codes that inform the progress through the tree are not fully understood,
but apparently the positional epigenetic information in the developing embryo
provides the information about the next development steps. In general terms,

\begin{quotation}
\noindent\lbrack t]hat model harks back to the \textquotedblleft developmental
landscape\textquotedblright\ proposed by Conrad Waddington in 1956. He likened
the process of a cell homing in on its fate to a ball rolling down a series of
ever-steepening valleys and forked paths. Cells had to acquire more and more
information to refine their positional knowledge over time --- as if zeroing
in on where and what they were through \textquotedblleft the 20 questions
game,\textquotedblright\ according to Jan\'{e} Kondev, a physicist at Brandeis
University. \cite{cepel:mathcells}
\end{quotation}

Again, the reference to the game of 20-questions reveals the common generative
mechanism of traversing a tree from the root to a specific leaf. 

\section{Selectionist and Generative Mechanisms}

There is a long tradition in biological thought of juxtaposing selectionism,
associated with Darwin, with instructionism, associated with Lamarck
(\cite{medawar:reith}; \cite{jerna:antibodies}). In an instructionist or
Lamarckian mechanism, the environment would transmit detailed instructions
about a certain adaptation to an organism, while in a selectionist mechanism,
a diverse variety of (perhaps random) variations would be generated, and then
some adaptations would be selected by the environment but without detailed
instructions from the environment. The discovery that the immune system was a
selectionist mechanism \cite{jerne:nat-sel} generated a wave of enthusiasm, a
\textquotedblleft Second Darwinian Revolution\textquotedblright%
\ \cite{cziko:womiracles}, for selectionist theories \cite{dennett:darwin}. 

In his Nobel Lecture \cite{jerne:nobel}, Niels Jerne even tried to draw
parallels between Chomsky's generative grammar and selectionism. One of the
distinctive features of a selectionist mechanism is that the possibilities
must be in some sense actualized or realized in order for selection to operate
on and differentially amplify or select some of the actual variants while the
others languish, atrophy, or die off. In the case of the human immune system,
"It is estimated that even in the absence of antigen stimulation a human makes
at least $10^{15}$ different antibody molecules---its preimmune antibody
repertoire." \cite[p. 1221]{alberts-et-al:mbio}. In Chomsky's critique of a
selectionist theory of universal grammar, he noted the computational
infeasibility of having representations of all possible human grammars in
order for linguistic experience and an evaluation criterion to perform a
selective function on them. The analysis of Chomsky's P\&P theory as a
generative mechanism instead suggests that the old juxtaposition of
\textquotedblleft selectionism versus instructionism\textquotedblright\ is not
the most useful framing for the study of biological mechanisms.

The discovery of the genetic code and DNA-RNA machinery for the production of
amino acids powerfully showed the existence of another biological mechanism, a
generative mechanism, that is quite distinct from a selectionist mechanism.
The examples of Chomsky's P\&P theory of grammar acquisition and the role of
stem cells in embryonic development provide more evidence of the importance of
generative mechanisms.

To better illustrate these two main candidates for biological mechanisms, it
might be useful to illustrate a selectionist and a generative mechanism in
solving the same problem of determining one among the $8=2^{3}$ options
considered in Figure 1. The eight possible outcomes might be represented as:
$|000\rangle,|100\rangle,|010\rangle,|110\rangle,|001\rangle,|101\rangle
,|011\rangle,|111\rangle.$

In the selectionist scheme, all eight variants are in some sense actualized or
realized in the initial state $S_{0}$ so that a fitness criterion or
evaluation metric (as in the FS scheme) can operate on them. Some variants do
better and some worse as indicated by the type size in Figure 7.
\begin{center}
\includegraphics[
height=2.0747in,
width=3.9597in
]%
{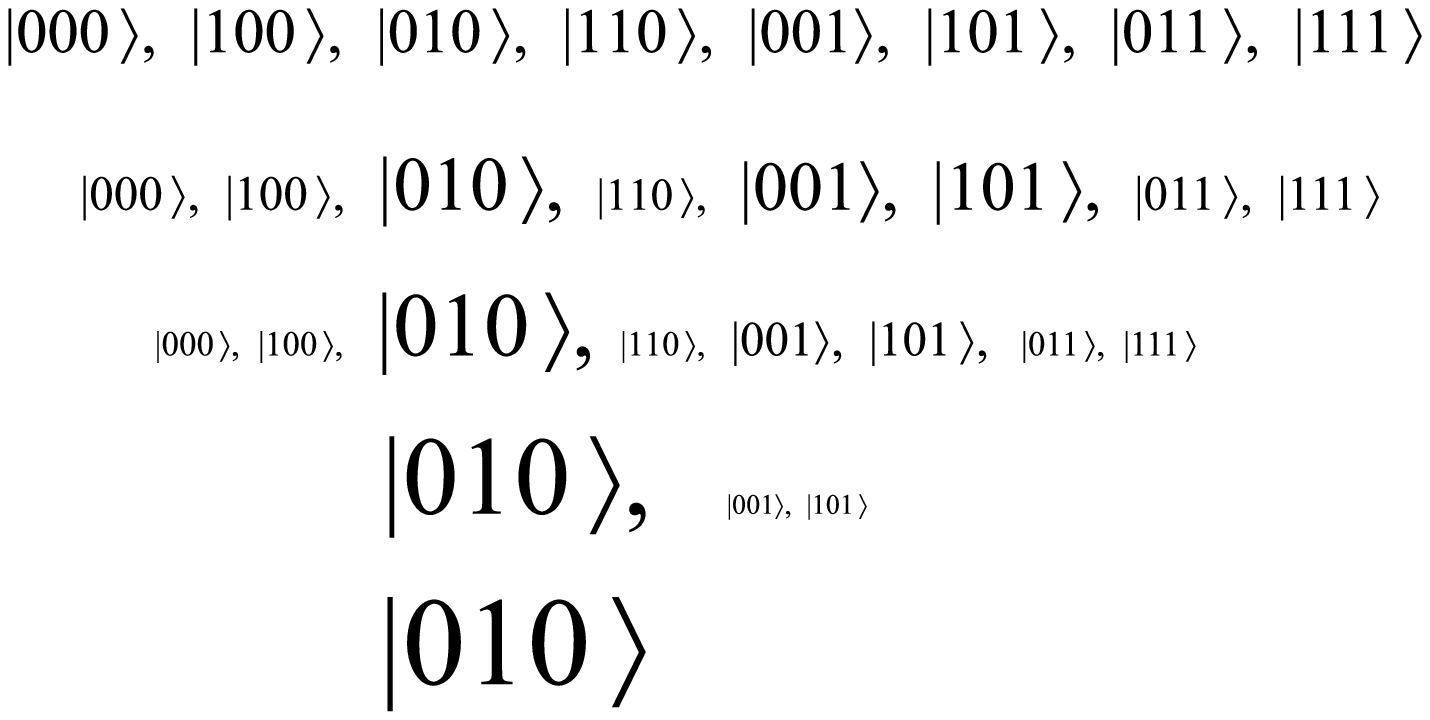}%
\end{center}

\begin{center}
Figure 7: A selectionist determination of the outcome $|010\rangle$.
\end{center}

\noindent The "unfit" options dwindle, atrophy, or die off leaving the most
fit option $|010\rangle$ as the final outcome.

With the generative mechanism, the initial state $S_{0}$ (the root of the
tree) is where all the switches are in neutral, so all the eight potential
outcomes are in a "superposition" (between left and right) state indicated by
the plus signs in the following Figure 8.
\begin{center}
\includegraphics[
height=1.7486in,
width=3.9597in
]%
{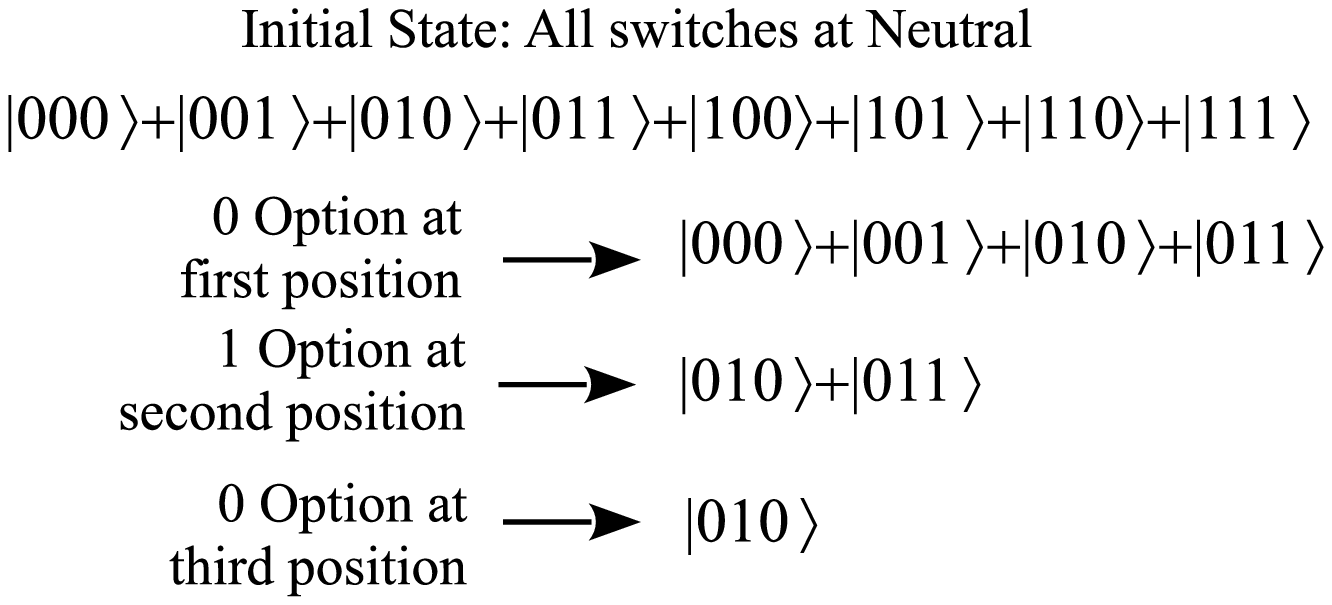}%
\end{center}

\begin{center}
Figure 8: A generative determination of the outcome $|010\rangle$.
\end{center}

The initial experience or first letter in the code sets the first switch to
the $0$ option which reduces the state to $|000\rangle+|001\rangle
+|010\rangle+|011\rangle$ (where the plus signs in the superposition of these
options indicate that the second and third switches are still in neutral).
Then subsequent experience sets the second switch to the $1$ option and the
third switch to the $0$ option. Thus, we reach the same outcome $|010\rangle$
as the final outcome in the two models but by quite different mechanisms. Note
that the generative mechanism `selects' a specific outcome but that does not
make it a `selectionist' mechanism.

There is another way to contrast a selective and generative mechanism. In
logic \cite{ell:intropartitions}, there are two principal lattices, the
lattice of subsets of a universe U and the lattice of partitions on $U$. For
$U=\{a,b,c\}$, the two lattices are pictured in Figure 9.
\begin{center}
\includegraphics[
height=1.8189in,
width=3.9597in
]%
{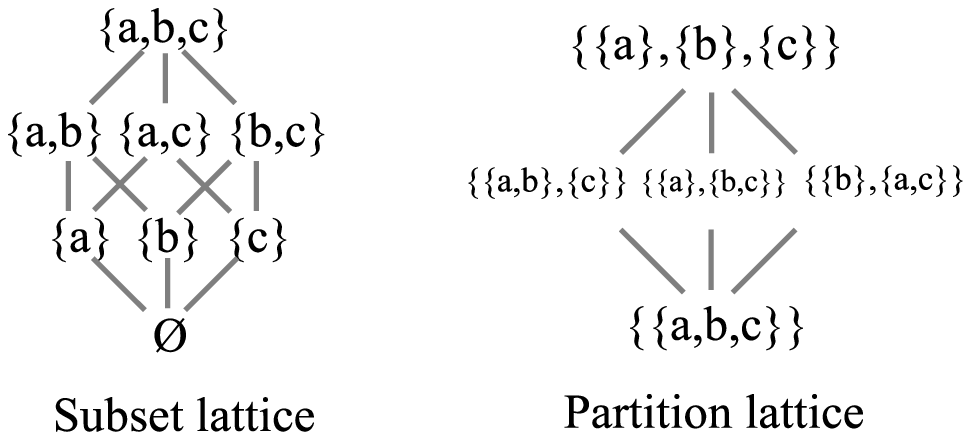}%
\end{center}

\begin{center}
Figure 9: The two basic logical lattices.
\end{center}

A selective mechanism to determine $a$, $b$, or $c$ from $U$ would start with
all the actual elements (top of the subset lattice) and then use a fitness or
evaluation criterion to narrow the set of actualities down to the selected
one. 

A generative mechanism starts with the indiscrete partition $\mathbf{0}%
_{U}=\{\{a,b,c\}\}=\{U\}$ (at the bottom of the partition lattice) where all
the elements are only potential outcomes not distinguished from each other.
Then distinctions are made, as represented by consecutive partition joins,
until an element is fully distinguished (i.e., appears as a singleton). A
coding scheme to determine each element is given in Table 2.

\begin{center}%
\begin{tabular}
[c]{|c|c|c|}\hline
Partitions & Consecutive Joins & Codes\\\hline\hline
$\{\{a,b,c\}\}=\mathbf{0}_{U}$ & $\{\{a,b,c\}\}=\mathbf{0}_{U}$ & \\\hline
$\{\{a\},\{b,c\}\}$ & $\{\{a\},\{b,c\}\}$ & $0=a$\\\hline
$\{\{a,b\},\{c\}\}$ & $\{a\},\{b\},\{c\}\}=\mathbf{1}_{U}$ & $10=b,11=c$%
\\\hline
\end{tabular}

Table 2: Coding scheme for $a$,$b$, or $c$.
\end{center}

The tree diagram for the Table 2 code is given in Figure 10.
\begin{center}
\includegraphics[
height=1.8782in,
width=2.5974in
]%
{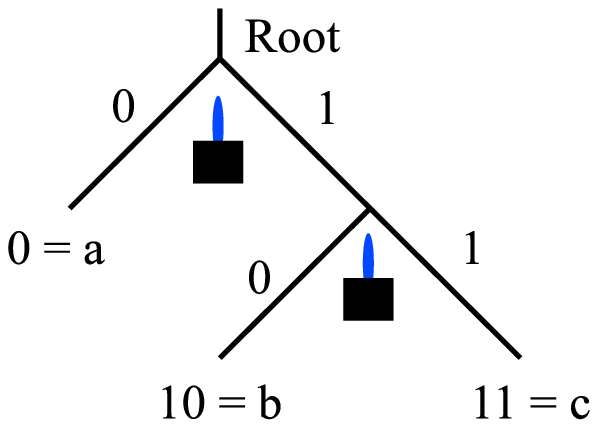}%
\end{center}

\begin{center}
Figure 10: Tree diagram for the code scheme of Table 2.
\end{center}

\section{Concluding remarks: The connection to information theory}

Finally, it should be mentioned that there is an intimate connection between
the tree diagrams representing generative mechanisms and information theory
(\cite{shannonweaver:comm}; \cite{ell:lit-igpl}). One can imagine a marble
rolling down from the root to one of the leaves where its path was
probabilistically determined at each branching point by a set of
probabilities. The simplest assumption on a binary tree is a half-half
probability of the marble taking each branch. From each leaf, there is a
unique path from the root to the leaf and the product of the probabilities
along that path gives the probability of reaching that leaf. With those
assumptions for the tree of Figure 1, each leaf has probability $1/2^{3}=1/8$.
Then the Shannon entropy of that probability distribution $p=(p_{1}%
,\ldots,p_{m})$ is: 

\begin{center}
$H(p)=\sum_{i}p_{i}\log_{2}(1/p_{i})=8\times1/8\times\log_{2}(1/(1/8))=\log
_{2}(2^{3})=3$
\end{center}

\noindent and the logical entropy is: 

\begin{center}
$h(p)=1-\sum_{i}p_{i}^{2}=1-8\times(1/8)^{2}=1-1/8=7/8.$
\end{center}

In this simple example, the Shannon entropy is the average number of binary
partitions (bits) needed to distinguish all the leaves on the tree
("messages"). The logical entropy always has the interpretation that on two
independent trials, there will be different outcomes. In this case, that is
the probability that on two independent rolls of a marble, it will end up at
different leaves. Since all the leaves are equiprobable, it is simply the
probability that the second marble took a different path (than on the first
trial), i.e., $1-1/8=7/8$.

With the same half-half branching probabilities for Figure 10, the leaf
probabilities are $\Pr(a)=1/2$ and $\Pr(b)=\Pr(c)=1/2\times1/2=1/4$. Then the
Shannon entropy is: 

\begin{center}
$H(p)=1/2\times\log_{2}(1/(1/2)+2\times1/4\times\log_{2}%
(1/(1/4))=1/2+(1/2)\times2=3/2$
\end{center}

\noindent which is, in this simple case, the average length of the code words
for the leaves. The logical entropy is: 

\begin{center}
$h(p)=1-(1/2)^{2}-2\times(1/4)^{2}=1-1/4-1/8=5/8$
\end{center}

\noindent which is always the probability that on two rolls, the marble will
end up at different leaves.

Tracing down a code tree from the root to the desired message generates the
code for that message on the sending side of a communications channel, and
then implementing the received code on the receiving side of the channel will
generate the received message. In the case of the biological examples (genetic
code, generative grammar, and embryonic development), the creation of the
codes is a matter of deep evolutionary history, so the focus of study is
usually on how those generative mechanisms implement codes to give specific outcomes.

\end{document}